\documentclass[a4paper,11pt]{article}
\usepackage{pos}

\renewcommand{\logo}{\relax}

\title{Fourier Acceleration in a Linear Sigma Model with Spontaneous Symmetry Breaking}
\ShortTitle{Fourier Acceleration in a Linear Sigma Model}

\author{Cameron Cianci}
\author{Luchang Jin}
\author*{Joshua Swaim}

\affiliation{University of Connecticut}

\emailAdd{joshua.swaim@uconn.edu}

\abstract{
	Fourier acceleration is a technique used in Hybrid Monte Carlo simulations to decrease the autocorrelation between subsequent field configurations in the generated ensemble. It has been shown, in the perturbative limit, to eliminate the problem of critical slowing down in a $\phi^4$ theory \cite{CSD_elim}. As a result, there are several techniques that are being explored to generalize Fourier acceleration to work with non-Abelian gauge theories like QCD \cite{RMHMC}\cite{GFFA}. It is hoped that these methods will prove effective at overcoming the problem of critical slowing down, even in the non-perturbative limit. In our work, we show that Fourier acceleration can be applied effectively to a linear sigma model in the symmetry broken phase, leading to reduced autocorrelation and faster thermalization. We present an algorithm for estimating the optimal Fourier acceleration masses dynamically, based on the lattice data. In the future, we hope to explore the effectiveness of these techniques in the strongly-interacting case. Since our $\phi^4$ theory is a linear chiral effective theory for QCD, this could be interesting for those who are seeking to generalize Fourier acceleration to QCD.
}

\FullConference{%
The 39th International Symposium on Lattice Field Theory,\\
8th-13th August, 2022,\\
Rheinische Friedrich-Wilhelms-Universität Bonn, Bonn, Germany
}

\begin{document}
\maketitle

\section{Introduction}
\subsection{The Linear Sigma Model}
We studied the linear sigma model in Euclidean spacetime with four scalar fields. The Lagrangian density for this theory is
$$\mathcal{L}(x) = \frac{1}{2}\sum_{\mu,i}(\partial_\mu\phi_i(x))^2+\frac{m^2}{2}\sum_i\phi_i(x)^2+\frac{\lambda}{4!}\left(\sum_i \phi_i(x)^2\right)^2+\alpha \phi_0(x),$$
where $i$ runs from 0 to 3. This model is of particular interest because it it can be used as a chiral effective theory for two-flavor QCD. When $\alpha=0$, this model has an $O(4)$ symmetry, which has the same Lie algebra as the chiral $SU(4)$ symmetry of QCD. The $\alpha\phi_0$ term that explicitly breaks this symmetry plays the role of the quark mass. When $m^2<0$, the symmetry is spontaneously broken, just like chiral symmetry in QCD. 

\subsection{The Hybrid Monte Carlo Algorithm}
To simulate this field theory, we used the Hybrid Monte Carlo method \cite{HMC}. We introduced a new set of four fields $\Pi_i(x)$ and simulated the classical field theory given by the Hamiltonian
$$H=\sum_{x}\left[\frac{1}{2M} \sum_i(\Pi_i(x))^2+\mathcal{L}(x)\right],$$
with each $\Pi_i(x)$ serving as the momentum conjugate to $\phi_i(x)$. Starting from some initial field configuration $\phi_i(x)$, we choose a random momentum field configuration $\Pi_i(x)$ according to the probability density $e^{-\Pi_i(x)^2/(2M)}$. Then we evolved $\phi_i(x)$ and $\Pi_i(x)$ using the classical equations of motion. After evolving for one unit of time, we performed a Metropolis accept/reject step to decide whether to add the new configuration $\phi_i(x)$ to our ensemble. Then the momentum field was updated again, and we repeated the process.

\subsection{Fourier Acceleration}
In our simulations, we employed a technique called Fourier acceleration to speed up our calculations \cite{FA}. We can write our Hamiltonian in momentum space as
$$H=\sum_{p}\left[\frac{1}{2M}\sum_i|\tilde\Pi_i(p)|^2+\tilde{\mathcal{L}}(p)\right].$$
We can modify our kinetic term by allowing $M$ to be different for different momentum modes.
$$H=\sum_{p}\left[\frac{1}{2}\sum_i \frac{|\tilde\Pi_i(p)|^2}{M_i(p)}+\tilde{\mathcal{L}}(p)\right].$$
In position space, this kinetic term can be written as $\frac{1}{2}\sum_{i,x,y}\Pi_i(x)G(x-y)\Pi_i(y)$ with the appropriate choice of the kernel $G(x-y)$. When we simulate using this new Hamiltonian, the rate at which each momentum mode evolves can be adjusted by adjusting $M_i(p)$. Now, in the free theory ($\lambda=\alpha=0$), each momentum mode evolves as an independent harmonic oscillator with angular frequency
$$\omega_{i}(p) = \sqrt\frac{m^2+8-2\sum_\mu\cos{p_\mu}}{M_i(p)},$$
($m$ is the mass parameter that appears in the Lagrangian, not the Fourier acceleration mass $M_i(p)$). We can choose $M_i(p)$ so that each mode evolves at the same rate, which allows us to avoid wasting time simulating rapid evolution for some modes while other modes evolve only a little. Even better, we can choose $M_i(p)$ so that $\omega_i(p)=\frac{\pi}{2}$ for each momentum mode. With this choice, the field configurations in our ensemble will be completely decorrelated. This is because each $\Pi_i(p)$ is $\pi/2$ radians out of phase with $\phi_i(p)$. Therefore, after evolving for one time unit with $\omega_i(p)=\pi/2$, the final field value $\phi_i(p)$ will be in phase with the initial momentum $\Pi_i(p)$, which was chosen randomly.

For non-free theories ($\lambda,\alpha\neq 0$), the momentum modes are not completely independent and do not behave like perfect harmonic oscillators. However, Fourier acceleration can still be used to decrease the correlation between subsequent field configurations in our ensemble.

\section{Methods}
\subsection{Choosing Fourier Acceleration Masses}
To determine the optimal Fourier acceleration masses to use for each mode, we used an iterative algorithm. For each trajectory, we calculated the average of the modulus of the real part of the force, $\langle |\text{Re}[\dot{\Pi}_i(p)]|\rangle$, on each momentum mode over the course of the HMC evolution. We also calculated the average of the modulus of the real part of the deviation of the field from its vacuum expectation value (which is zero for all modes except the zero mode of $\phi_0$) $\langle |\text{Re}[\phi_i(p)-\sigma_{vev}\delta_{i,0}\delta_{p,0}]|\rangle$. If we make the assumption that each mode is an independent harmonic oscillator, then 
$$\omega_i^2(p)=\frac{1}{M_i(p)}\frac{\langle |\dot{\Pi}_i(p)|\rangle}{\langle |\phi_i(p)-\sigma_{vev}\delta_{i,0}\delta_{p,0}|\rangle}.$$
Our basic algorithm works as follows:
\begin{itemize}
	\item For a given set of Fourier acceleration masses $M_i(p)$, run a batch of trajectories. On each trajectory, calculate
	$\langle |\dot{\Pi}_i(p)|\rangle$ and $\langle |\phi_i(p)-\sigma_{vev}\delta_{i,0}\delta_{p,0}|\rangle.$
	\item Calculate the average of $\langle |\dot{\Pi}_i(p)|\rangle$ and $\langle |\phi_i(p)-\sigma_{vev}\delta_{i,0}\delta_{p,0}|\rangle$ over the preceding batch of trajectories and take their ratio $\frac{\langle |\dot{\Pi}_i(p)|\rangle}{\langle |\phi_i(p)-\sigma_{vev}\delta_{i,0}\delta_{p,0}|\rangle}.$
	\item Choose a new set of Fourier acceleration masses by taking $$M_i(p)=\frac{4}{\pi^2}\frac{\langle |\dot{\Pi}_i(p)|\rangle}{\langle |\phi_i(p)-\sigma_{vev}\delta_{i,0}\delta_{p,0}|\rangle}.$$
	This choice is designed with the goal of making each mode oscillate at angular frequency $\omega=\frac{\pi}{2}$.
	\item To avoid numerical instabilities, impose a lower bound on the masses proportional to $\langle |\dot\Pi_i(p)|\rangle$, with the constant of proportionality set by hand as appropriate for different parameters.
	\item Impose an absolute lower bound on the masses, also set by hand.
	\item Repeat this procedure, using the new Fourier acceleration masses, for a new batch of trajectories.
\end{itemize}

\subsection{Measuring Observables}
\subsubsection{Particle Masses}
The term $\alpha \phi_0$ in the Lagrangian picks out a unique ground state so that when the $O(4)$ symmetry is spontaneously broken, the $\phi_i$ degrees of freedom for $i=1,2,3$ always correspond to the pion degrees of freedom (the approximate goldstone bosons that result from the spontaneous breaking of the $O(4)$ symmetry). Meanwhile, the $\phi_0$ degree of freedom, minus its vacuum expectation value, corresponds to the sigma degree of freedom. We calculate the pion and sigma masses by fitting the correlation functions of these fields.

\subsubsection{The Pion Decay Constant}
To determine the effective pion decay constant $F_\pi$, we start with the continuum relation
$$\langle 0|A_0^i(\mathbf{x},t)|\pi^i(\mathbf{p}=0)\rangle=F_\pi m_\pi e^{-m_\pi t}.$$
On the finite lattice, this becomes
$$\langle0|A_0^i(\mathbf{x},t)|\pi^i(\mathbf{p}=0)\rangle=2F_\pi m_\pi e^{-m_\pi N_T/2}\sinh((N_T/2-t+1/2)m_\pi),$$
where $A^i_\mu(x)=\phi_0(x-\mu)\phi_i(x)-\phi_0(x)\phi_i(x-\mu)$ is a discretization of the conserved current associated with the broken part of the $SO(4)$ symmetry. We use $t-1/2$ instead of $t$ in the above equation because our lattice definition of $A_0^i$ combines fields at $t$ and $t-1$.
In terms of the $\phi_i$ fields, this gives us
$$F_\pi = \frac{\langle0|\left(\sum_\mathbf{x}A_0^i(\mathbf{x},t)\right)\left(\sum_{\mathbf{x}'}\phi_i(\mathbf{x}',0)\right)|0\rangle}{\sqrt{\langle0| \left(\sum_{\mathbf{x}'}\phi_i(\mathbf{x}',t')\right)\left(\sum_{\mathbf{x}''}\phi_i(\mathbf{x}'',0)\right)|0\rangle}}\frac{\sqrt{\cosh((N_T/2-t')m_\pi)}}{\sqrt{m_\pi V_xe^{-m_\pi N_T/2}}\sinh((N_T/2-t+1/2)m_\pi)}.$$

We could just extract the decay constant using the above expression, but to get a more stable fit, we can use Noether's theorem for the lattice to get $\sum_\mu \big(A_\mu^i(x+\mu)-A_\mu^i(x)\big)=-\alpha\phi^i(x).$ From this, we get
$$F_\pi=-\frac{\alpha \langle 0| \phi^i(x)|{\pi^i(\mathbf{p}=0)}\rangle}{2m_\pi e^{-N_Tm_\pi/2}\big(\sinh((N_T/2-t-1/2)m_\pi)-\sinh((N_T/2-t+1/2)m_\pi)\big)}.$$
In terms of the $\phi_i$ fields, this is
$$F_\pi=-\frac{\alpha\sqrt{\cosh((N_T/2-t)m_\pi)} \sqrt{\langle 0| \left(\sum_\mathbf{x}\phi^i(\mathbf{x},t)\right)\left(\sum_{\mathbf{x}'}\phi_i(\mathbf{x}',0)\right)|0\rangle}}{\sqrt{m_\pi V_x} e^{-N_Tm_\pi/4}\big(\sinh((N_T/2-t-1/2)m_\pi)-\sinh((N_T/2-t+1/2)m_\pi)\big)}.$$
We got the cleanest results by using this equation to solve for $\sqrt{\langle 0| \left(\sum_\mathbf{x}\phi^i(\mathbf{x},t)\right)\left(\sum_{\mathbf{x}'}\phi_i(\mathbf{x}',0)\right)|0\rangle}$ and combining it with our earlier equation for $F_\pi$.



\section{Results}
\begin{figure}
	\centering
	\includegraphics[width=\textwidth]{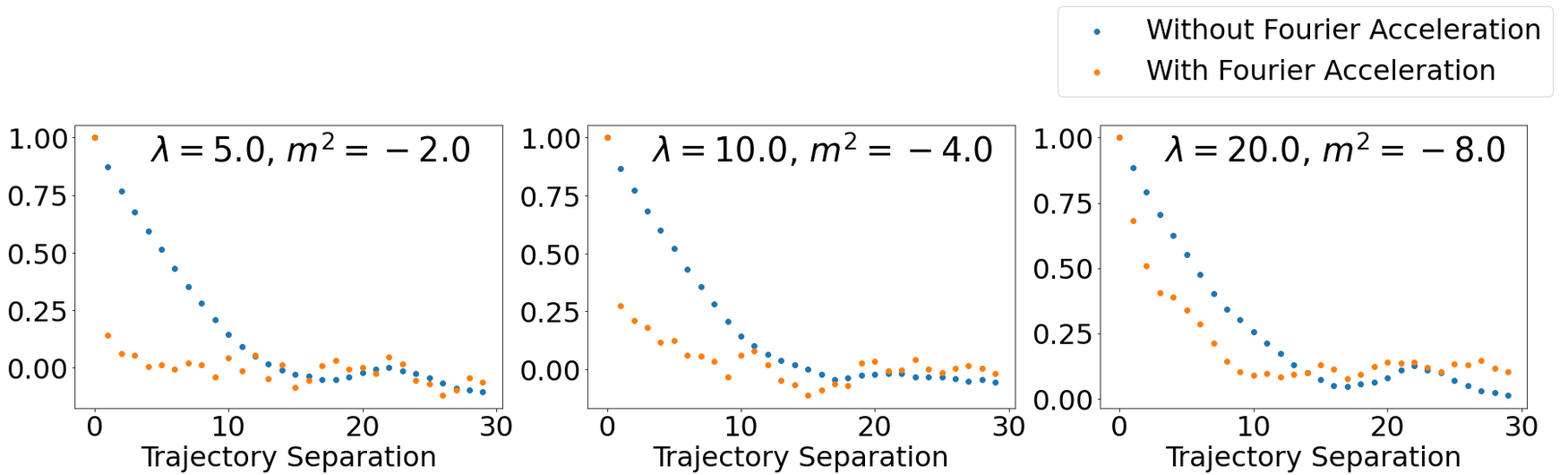}
	\caption{The autocorrelation of pion mass jackknife samples on an $8^3\times 16$ lattice, $\alpha=0.1$. For each of these simulations, $m_\pi\approx 0.3$ and $F_\pi\approx 1.4\pm 0.1$. From left to right, $m_\sigma\approx 1.0$, 1.3 and 1.5 respectively.}
	\label{autocorr}
\end{figure}

\begin{figure}
	\centering
	\includegraphics[width=\linewidth]{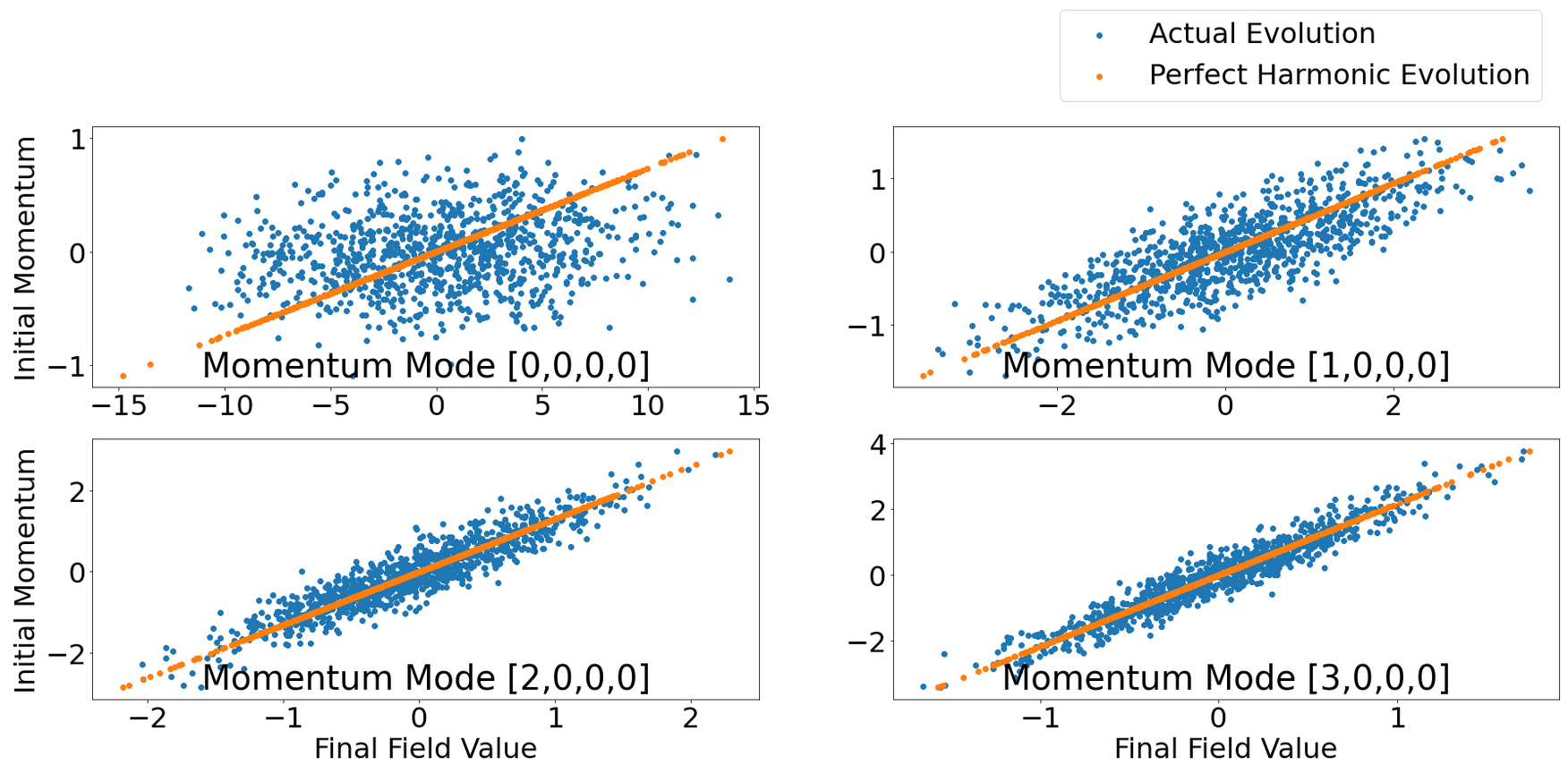}
	\caption{The real part of the pion field at end of a trajectory versus the real part of the momentum at the start of the trajectory, at four different sites in momentum space. For perfect harmonic evolution, the initial momentum is in phase with the final field value: $\text{Re}(\Pi_{i,\text{initial}}(p))=-M\omega^2\text{Re}(\phi_{i,\text{final}}(0))$. This simulation was done on an $8^3\times 16$ lattice with $m^2=-4.0$, $\lambda=10.0$, $\alpha=0.1$. Observable values were $m_\pi\approx 0.309\pm0.005$, $m_\sigma\approx 1.26\pm0.02$ and $F_\pi\approx 1.43\pm0.02$}.
	\label{evolution}
\end{figure}

We found that Fourier acceleration can be effective at reducing the autocorrelation length of observables like the pion mass. However, it became less and less effective for larger $\lambda$ (see Figure \ref{autocorr}). As indicated in Figure \ref{evolution}, the assumption that each mode evolved as an independent harmonic oscillator was approximately valid, especially for high momentum modes, but became less and less valid as $\lambda$ increased.


\section{Conclusion}
We have seen the Fourier acceleration can be very beneficial for simulating the linear sigma model with spontaneous symmetry breaking, as long as the interaction strength $\lambda$ is not too large. Unfortunately, simulating this effective field theory at a physical ratio of pion mass to pion decay constant requires setting $\lambda$ to be very large. 

Since our implementation of Fourier acceleration relies on the field values at each point in momentum space being approximately independent, it makes sense that the method breaks down for larger $\lambda$. In the limit as $\lambda$ goes to infinity, the field will obey the constraint that $\sum_i(\phi_i(x))^2$ becomes a constant. Since this is a set of local constraints in position space, it becomes a set of non-local constraints in Fourier space that force the momentum modes to be highly correlated. In the future, we intend to explore other if our methods can be generalized for use in theories with large $\lambda$.

\bibliography{template_lat22}

\end{document}